# OCCURENCE OF ACTIVE LAYER OPTICAL PROPERTIES ON SOME LASING CHARACTERISTICS DYE-DOPED Ch LC

I. Ilchishin[1] [*], Eu. Tikhonov[1], V. Belyakov[2]

[1] Institute of Physics, National Academy of Sciences, Nauky prosp. 46, 03028, Kyiv, Ukraine
[2] L.D.Landau Institute for Theoretical Physics, Kosygin str.2, 119334, Moscow, Russia
[*] Corresponding author: AN e-mail address: lclas@iop.kiev.ua
   phone: +38-044 5250929, fax: +38-044 5251589

**Abstract**

Effect of a planar texture quality and its thickness on lasing spectrums and thresholds in dye-doped cholesteric liquid crystal (CLC) of steroid type is explored. Transition from the qualitative planar texture to the poor texture quality is accompanied by change of characteristic mode structures and by shift of barycentre in the long-wave side and the considerable growth of the lasing threshold. It is found that in the CLC texture created by substrates with perpendicular directions of orientation the stable single-mode lasing takes place. The nature of oscillated modes in such texture is caused by phase jump. The gained results show that in steroid CLC, unlike induced one, lasing spectra is possible to feature with the coupled wave model.

## 1. Introduction

In spectrums of dye-doped CLC lasers the different values of an oscillation frequency concerning of selective reflection (SR) band are registered experimentally. Even in early work [1] it has been established that in CLC-materials with small birefringence what are steroid CLC (cholesterol derivatives), the lasing spectrum is located near to centre of a SR band and its performances can be explained within the limits of coupled wave model in medium with the space modulation of refraction index [2]. For a standing of oscillation spectrum on the SR band edge that is observed on induced CLC [1] with strong birefringence, for the first time the univocal explanation was abcent because the big modulation refraction index went outside the limits applicability of coupled wave model.

Thereupon authors of publication [3] have advanced a hypothesis about the lasing spectrum standing on the edge of a SB band CLC by analogy to the predicted lasing spectrum standing near edge of the photonic band gap for the photonic crystals [4]. Therefore the possible parent on which oscillation spectrum from dye doped CLC and also their some other performances can explain within the limits of photonic crystal model. In this approach the standing of lasing spectrum on centre of the SB band connect with inside defect of the photonic crystal and modes of the structured defects [5]. The analogous approach concerning the nature of lasing spectrum with participation of modes of the defective structure is used also in the fresh theoretical model [6].

In connection with the our purpose the study was directed on revealing of the steroid type CLC optical quality on spectroscopic and threshold performances of lasing. Our purpose also included the determination of applicability of existing theoretical models to lasing behavior of CLC with small birefringence ($\Delta n \approx 0,05$).

## 2. Experimental performance

In the capacity of the active textures CLC 3-component mixture of oleate, pelargonate and cholesterol chloride with temperature modification of spiral pitch ≈3 nm/grade were used. CLC was doped benzanthrone or phenolenone dyes at weight concentration of 0,2-0,3 %. The maximum of the SR band was chosen to be superimposed with a fluorescence maximum in region ≈600нм. The planar texture was produced by the known method of a surface roll forming by rubbing substrates and their additional shift as it was featured in [7]. Optical pumping of the active textures of CLC was carried out $2^{nd}$ harmonic (530 nm) $Nd^{3+}$ laser operating in seldom repeated pulse mode with pulse duration $\cong$20ns. The lasing spectrum of dyed CLC matching to each pumping pulse were mapped in a focal plane of a spectrograph with an inverse dispersion 0,6 nm/mm and displaid by the web camera on PC.

## 3. Results and discussion

The spectra and thresholds of laser oscillation were explored at different thicknesses of the active films and some expedients of a planar texture making. It was established that for qualitative on spectroscopic parameters (halfwidth of SR band, level of scattering) planar texture, the lasing spectrum consists of 3 narrow lines matching lowest longitudinal modes of the DFB- laser at rather low lasing threshold (40-60кВт/см$^2$). On Fig. 1. transmission spectra of planar texture of the steroid CLC doped phenolenone dye № 490 (≈0,3 % on weight, absorption peak 540 nm) are presented. Presence of the transparent electrodes $SnO_2$ led to appreciable improvement of planar texture: the SR band narrows more than on 10 % in comparison with an analogous SR band for texture with an orienting layer of polyimide lacquer (a curve 2, Fig. 1.). Besides intensity of diffraction Bragg maximum (a curve 1 Fig. 1) in the region 600 nm considerably increased).

The standing of lasing spectra (an arrow on Fig. 1) shows that in the case of better planar texture, the lasing spectrum barycentre practically coincides (difference <1nm,) with centre of the SR band. Deterioration of texture which is observed in the absence of $SnO_2$ layer , apparently from a curve 2 on Fig. 1., shifts of lasing spectrum towards a right edge of the SR band.

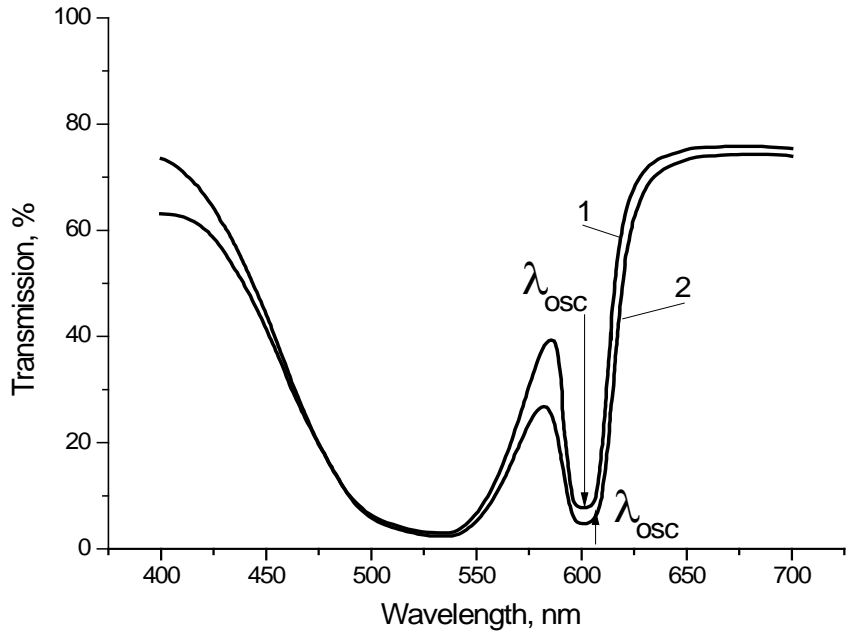

**Figure 1**

On Fig. 2 the lasing spectra of the dyed CLC sample which transmission spectrum is seen on Fig. 1 ( curve 1) as function of pumping power are presented. Apparently for the qualitative planar texture the lasing spectrum is disposed on centre of SR band and characterized by rather narrow lines (longitudinal modes). Growth of pumping the number of observable modes remains invariable up to region of sample fracture at pumping> 430 $I_{пор.}$. Presence only 3 modes in the lasing spectrum in a wide limits of pumping testifies on the strong threshold mode selections and are agreed with the theory of helical DFB- laser , developed in works [8-9]. Meanwhile lasing spectrum of CLC with major birefringence are characterized more than 3 modes as it was observed by authors [3].

The lasing spectrum for the sample which transmission is featured by dependence 2 on Fig. 1, is presented on Fig.3. Besides the spectrum was shifted in the long-wave side SR band, the deterioration of planar texture has led to an essential broadening of a mode spectrum and the strong growth of the lasing threshold. Comparing the data Fig. 1, 2 and 3 we come to conclusion that at almost equal absorption on the given pumping, thickness of both samples 45μm, the better optical quality planar texture with layer $SnO_2$ has the lasing threshold on two orders lower.

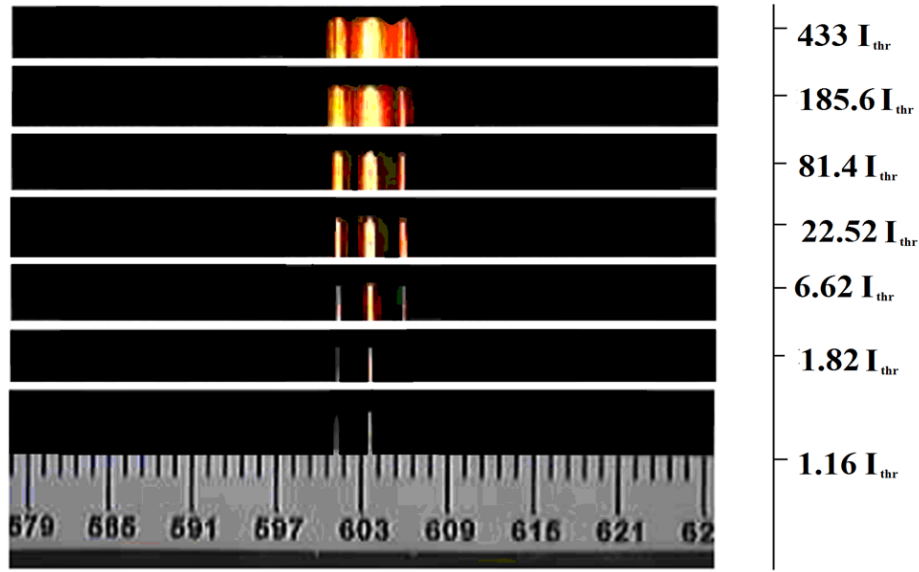

**Figure 2**

Similar results are observed when the active film thickness changes within limits (45-160) μm. For qualitative samples CLC with $SnO_2$ on substrates the thickness magnification leads gradual broadening of lasing mode spectrum and moving the spectrum barycentre to the long-wave edge of its SR band.

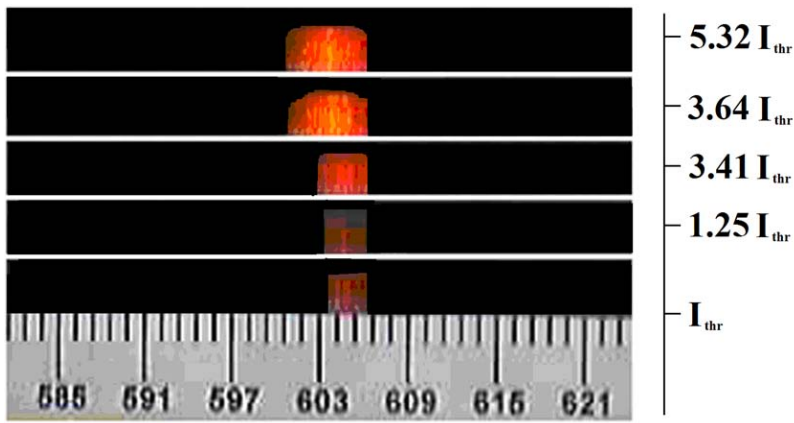

**Figure 3**

Important results have been gained in the case of active planar textures CLC prepared by rubbing of orienting substrates in two crossly perpendicular directions. In such textures, apparently from Fig. 4., there is the stronger threshold selection of modes so that oscillation arises on the principal mode with the lowest threshold of excitation only.

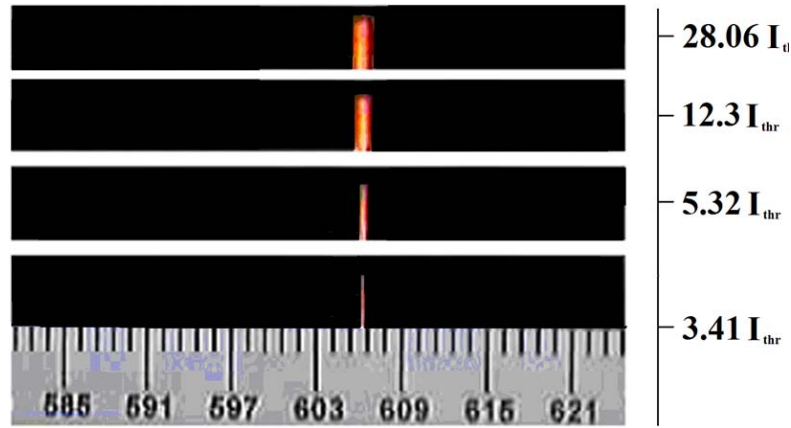

**Figure 4**

Apparently from Fig. 4. pumping growth leads to only visible (hardware) broadening of this line, as well as a case on Fig. 2. The obtained data testifies to an essential decrease (on orders) lasing line width for the case. More precise measurements about the line width will be presented in separate work with use interferometer technique.

Possible explanation of above mentioned lasing behaviour is formation of local "defect mode" due to a phase jump in texture at the mismatched turn of oriented substrates. The following lasing behaviour on "defect mode" as a phase jump was described with the work [6] considered as defect an insertion of an isotropic layer into a perfect cholesteric structure. Main expressions of the work are directly applicable to the our case of a phase jump of the cholesteric spiral if the factor „**kd**", where **k** is the wave vector and **d** is the inserted isotropic layer thickness, is substituted in the formulas of the work [6] by $\Delta\varphi$, as the phase jump of the cholesteric spiral.

The corresponding lasing threshold for the defect mode frequency at the centre of stop band reaches the lowest value for the phase jump angle $\Delta\varphi = 90°$. To clarify the origin of the observed effect it is desirable to check some other theoretical predictions related to the defect modes, namely, the connection of the phase jump and position of the defect mode frequency inside the stop band. Rotation of the plate by an angle less than 90° in the experiment has to results in approaching the defect mode frequency (i.e. the lasing frequency) to the high frequency edge of stop band and, correspondent ly, the rotation of the plate by an angle larger than 90° result in approaching the defect mode frequency (i.e. the lasing frequency) to the low frequency edge of the stop band. In both cases (rotation angle $\Leftrightarrow$ 90°) the lasing threshold should be higher than for the rotation angle 90°. In the general case the value $\gamma$ (lasing threshold value of the gain normalised by the light wave vector) has to be found by numerical approach. However for the case of thick cholesteric layers of thickness L (such that the condition $|qL| \gg 1$ is fullfieled, where **q = to {1 + τ/2 κ) 2 - [(τ / to) 2 + δ2]** $^{1/2}$**}** $^{1/2}$, δ is the cholesteric dielectric anisotropy and $\tau = 4\pi/p$, where p is the cholesteric pitch) an analytic expression for the dependence of $\gamma$ on the plate rotation angle (the phase jump angle $\Delta\varphi$) may be found. The corresponding expression for the value of $\gamma$ if $|qL| \gg 1$ at phase jump $\Delta\varphi = 90°$ is given by the following formula:

$$\gamma = [4p / (3\pi L)] \exp[-\delta\pi L/p] \qquad (1)$$

The dependence of lasing threshold $\gamma$ on the plate rotation angle (the phase jump angle $\Delta\varphi$) in the angular range where the condition $|qL| \gg 1$ is fullfieled is presented at fig.5. The calculation at the fig.5 performed for the CLC layer thickness 40 spiral halfturns and $\delta = 0.05$.

We did not manage still to register a appropriate dip in SRB which existence is predicted by the theory [6] at generation with participation of the defective texture. Possible reason of failure - the typical method of sample transmission measuring which uses big squares of the sample while transmission dip can be connects with a small local region of sample.

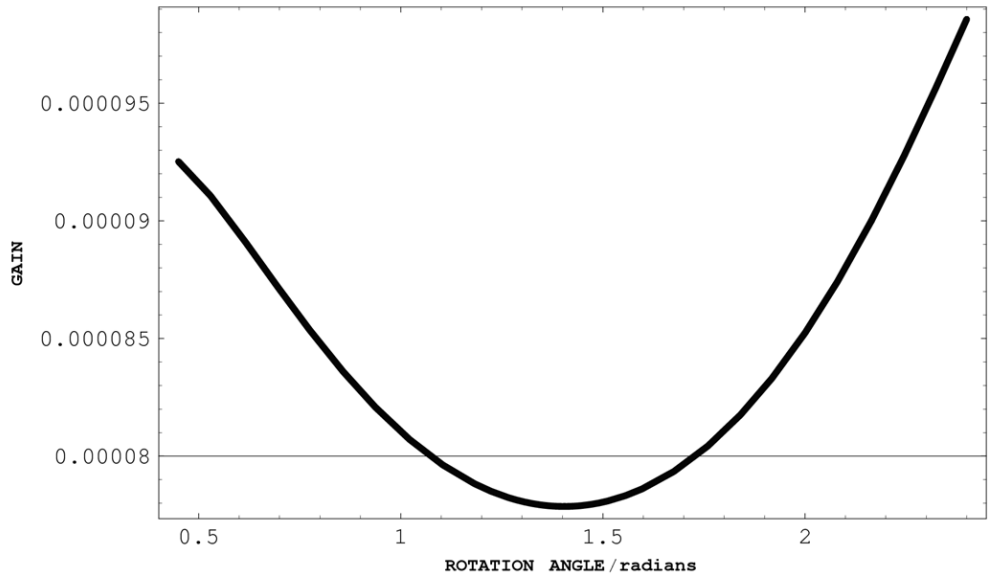

**Figure 5**

### 4. Conclusions

1. In the lasing sample based on derivative cholesterol with small birefringence unlike induced ChLC, the lasing spectrum is disposed nearby the centre of SR band. Than narrower is the SR band and than narrower are the lasing lines, the nearer to the centre of SR band the lasing lines are placed. It takes place at the use of substrates with $SnO_2$ layer. Use of substrates without $SnO_2$ layer results in broadening of SR band up to zero of diffraction reflection, disappearance of sharp linear mode structure in a lasing spectrum and its displacement to the right edge of SR band.

2. Similar displacament of lasing spectra from the centre of SR band to its edge, resulted at optical worsening of quality of planar texture, is observed at the increase of its thickness, accompanied by worsening of spiral texture ordering. It contradicts with possibility of explanation lasing on the centre of SR band as lasing on "defect mode" and as unhinge-quality planar structure. At a such explanation lasing spectra with film thickness growth should be moved to the centre of SR band.

3. Single mode lasing at the crossly oriented substrates can be related to the presence of phase jump in the spiral structure of CLC. Existing theoretical conception [4,6] allow to interpret the single mode lasing at the turn of orienting direction on one of substrates as display of imperfect mode lasing, although the transmission spectrum expected dip does not discover.

4. Descriptions of DFB-lasing in dye doped planar textures prepared on derivative cholesterol corresponds to theoretical models [2,8-9], and DFB-lasing - prepared on induced CLC has better correspondence with models [4, 6].

**Acknowledgment**: This work was supported by the RFBR grant 09-02-90417-Ukr_f_a and by the FRSF grant F 28.2/084 (Ukr-Russ).

## References

1. I.P. Ilchishin, Eu. A.Tikhonov, A.V.Tolmachev et. al. *Mol. Cryst. Liq. Cryst.* 191,351 (1990).

2. H.Kogelnik, and S.V. Shank. *J. Appl. Phys.* **43**, 2327 (1972).

3 I. Kopp, B. Fan, H. K. M. Vthana, and A. Z. Genack, *Opt. Lett*. **23**, 1707 (1998).

4. J. P. Dowling, M. Scalora, M. J. Bloemer, C. M. Bowden, *J. Appl. Phys*. **75**, 1896 (1994).

5. R.Ozaki, T.Matsui, M.Ozaki, K. Yoshino. *Appl. Phys. Lett*. **82**, 3593 (2003).

6. V.A. Belyakov. *Mol. Cryst. Liq. Cryst*. **494**, 127 (2008).

7. Yu V. Denisov, V.A. Kizel, and E.P. Sukhenko. *JETP,* **71**, 679 (1976).

8. F.K.Kneubuhl. *Infrared Physics*. **23**, 115 (1983).

9. H.P.Preiswerk, M.Lubanski, S. Gnepf, and F.K.Kneubuhl. *IEEE J*. **QE-19**, 1452 (1983).

## Figure captions

**Figure 1** Transmission spectra of phenalenone dye in the mixture of CLC next composition: 40 % cholesteryl oleate, 35 % cholesteryl pelargonate and 25 % cholesteryl chloride.

1- glass substrates with $SnO_2$ and polyimide lacquer layers; 2 - glass substrates with polyimide lacquer layer. The layer thickness is 45 μm. Arrows positions of lasing spectra are indicated.

**Figure 2**.The lasing spectra as pumping intensity for a sample, that indicated on a Fig.1 with a layer

SnO$_2$ on substrates (transmission spectrum 1). High mode selection is the number of longitudinal modes in a spectrum does not change with pumping increase. The layer thickness is 45 μm.

**Figure 3**. The lasing spectra as pumping intensity for sample, indicated on a Fig.1 without a layer SnO$_2$ on substrates (transmission spectrum 2 on Fig.1). A mode structure disappears and lasing thresholds increases on two orders. The layer thickness is 45 μm.

**Figure 4**. The lasing spectra as pumping intensity for a sample, indicated on a Fig.1 with turn on 90º one of orienting substrates. The layer thickness is 45 μm. The single mode lasing is saved up to the high levels of pumping, that results in destruction of sample.

**Figure 5.** The dependence of lasing threshold γ on the plate rotation angle where the condition |qL| »1 is fullfieled (calculation).